\documentclass[twocolumn]{aastex62}

\usepackage{bm}

\graphicspath{./}

\received{2018 May 11}
\accepted{2018 May 18}
\published{2018 June 6}
\submitjournal{ApJL}

\begin{document}

\title{Measuring the viewing angle of GW170817 with electromagnetic and gravitational waves}

\correspondingauthor{Daniel Finstad}
\email{dfinstad@syr.edu}

\author[0000-0002-8133-3557]{Daniel Finstad}
\affil{Department of Physics, Syracuse University, Syracuse, NY 13244, USA}

\author[0000-0002-3316-5149]{Soumi De}
\affiliation{Department of Physics, Syracuse University, Syracuse, NY 13244, USA}

\author[0000-0002-9180-5765]{Duncan A. Brown}
\affiliation{Department of Physics, Syracuse University, Syracuse, NY 13244, USA}

\author[0000-0002-9392-9681]{Edo Berger}
\affiliation{Harvard-Smithsonian Center for Astrophysics, 60 Garden Street, Cambridge, MA 02139, USA}

\author[0000-0002-5946-2467]{Christopher M. Biwer}
\affiliation{Department of Physics, Syracuse University, Syracuse, NY 13244, USA}
\affiliation{Applied Computer Science (CCS-7), Los Alamos National Laboratory, Los Alamos, NM, 87545, USA}



\begin{abstract}

The joint detection of gravitational waves (GWs) and electromagnetic (EM) radiation from the binary neutron star merger GW170817 ushered in a new era of multi-messenger astronomy. Joint GW--EM observations can be used to measure the parameters of the binary with better precision than either observation alone. Here, we use joint GW--EM observations to measure the viewing angle of GW170817, the angle between the binary's angular momentum and the line of sight. We combine a direct measurement of the distance to the host galaxy of GW170817 (NGC\,4993) of $40.7\pm 2.36$ Mpc with the Laser Interferometer Gravitational-wave Observatory (LIGO)/Virgo GW data and find that the viewing angle is $32^{+10}_{-13}\,\pm 1.7$ degrees (90\% confidence, statistical, and systematic errors). We place a conservative lower limit on the viewing angle of $\ge 13^\circ$, which is robust to the choice of prior. This measurement provides a constraint on models of the prompt $\gamma$-ray and radio/X-ray afterglow emission associated with the merger; for example, it is consistent with the off-axis viewing angle inferred for a structured jet model. We provide for the first time the full posterior samples from Bayesian parameter estimation of LIGO/Virgo data to enable further analysis by the community.

\end{abstract}

\keywords{binaries---close, stars---neutron, gravitational waves}


\section{Introduction} \label{sec:intro}

On 2017 August 17, the Advanced Laser Interferometer Gravitational-wave Observatory (LIGO) and Virgo observed the gravitational waves (GWs) from a binary neutron star merger, dubbed GW170817 \citep{TheLIGOScientific:2017qsa}. This signal was followed $1.7$ s later by a short gamma-ray burst (GRB), GRB170817A, detected by the {\it Fermi} and {\it INTEGRAL} satellites \citep{Goldstein:2017mmi,Savchenko:2017ffs}. Rapid follow-up of the LIGO/Virgo sky localization region led to the identification of an optical counterpart in the galaxy NGC\,4993 \citep{Coulter:2017wya,Soares-Santos:2017lru,Valenti:2017ngx}, which in turn enabled multi-wavelength observations spanning from radio to X-rays. 

Ultraviolet, optical, and near-infrared observations covering the first month post-merger led to the inference of a complex ejecta structure in terms of mass, velocity, and opacity (e.g., \citealt{Chornock:2017sdf,Cowperthwaite:2017dyu,Kasliwal:2017ngb,Nicholl:2017ahq,Pian:2017gtc,Smartt:2017fuw,Villar:2017wcc}), potentially indicative of non-spherical angular structure.  Radio and X-ray observations revealed brightening emission for the first $\approx 5$ months, which has been interpreted as resulting from an off-axis structured relativistic jet (e.g., \citealt{Alexander:2017aly,Alexander2018,Lazzati:2017zsj,Margutti:2017cjl,Margutti:2018xqd}), or alternatively a spherical ``cocoon'' of mildly relativistic ejecta (e.g., \citealt{Mooley:2017enz}).

Measuring the angle between the binary's angular momentum axis and the line of sight is important for an understanding of the engine powering the multi-wavelength electromagnetic (EM) emission from GW170817.  Following \cite{TheLIGOScientific:2017qsa}, we define the viewing angle  $\Theta=\min(\theta_{JN},180^\circ - \theta_{JN})$, where $\theta_{JN}$ is the angle between the binary's total angular momentum and the line of sight \citep{TheLIGOScientific:2017qsa}. For systems where the angular momentum of each compact object (the spin) is small, and precession of the binary's orbital plane is not significant (as is the case for GW170817), $\theta_{JN} \approx \iota$, where $\iota$ is the angle between the binary's orbital angular momentum and the line of sight (the inclination angle). There is a degeneracy between the binary's inclination, $\iota$, and the luminosity distance, $d_L$, when only LIGO/Virgo observations are used to measure the inclination angle \citep{Wahlquist:1987rx}. Breaking this degeneracy with an independent distance measurement immediately allows one to place tighter constraints on the inclination angle \citep{Fan:2014kka}.

Using GW observations alone, LIGO and Virgo constrained the viewing angle to $\Theta \le 55^\circ$ at 90\% confidence with a low-spin prior \citep{TheLIGOScientific:2017qsa}. To provide an independent distance measurement, Abbott {\it et al.} used the estimated Hubble flow velocity for NGC\,4993 of $3017\pm 166$ km s$^{-1}$ and a flat cosmology with $H_0=67.90\pm 0.55$ km s$^{-1}$ Mpc$^{-1}$ to constrain $\Theta\le 28^\circ$\citep{TheLIGOScientific:2017qsa}. \cite{Mandel:2017fwk} used the combined $H_0$-inclination posterior from \cite{Abbott:2017xzu} in conjunction with the Dark Energy Survey measurement of $H_0=67.2^{+1.2}_{-1.0}$ km s$^{-1}$ Mpc$^{-1}$ \citep{Abbott:2017smn} to infer $\Theta\le 28^\circ$ at 90\% confidence \citep{Mandel:2017fwk}. These circuitous approaches to breaking the distance-inclination degeneracy were motivated partly by the absence of a precise distance measurement to NGC\,4993, as well as by the lack of a published distance-inclination posterior probability distribution. Furthermore, \cite{Mandel:2017fwk} was not able to place a strong constraint on the lower bound of $\Theta$, as his analysis used the GW posteriors \citep{Mandel:2017fwk} and was constrained by LIGO/Virgo's choice of prior in their GW analysis \citep{Abbott:2017xzu}. 

Here, we directly use the most precise distance measurement available for NGC\,4993 of $d_L=40.7\pm 2.36$ Mpc \citep{Cantiello:2018ffy} and the LIGO/Virgo GW data \citep{TheLIGOScientific:2017qsa} to infer $\Theta$ directly from joint GW--EM observations using Bayesian parameter estimation \citep{emcee,pycbc-inference,pycbc-software}. To allow our results to be used by the community for further analysis we provide the full posterior samples from our analysis as supplemental materials.

\section{Methods} \label{sec:methods}

We use Bayesian inference to measure the parameters of GW170817 \citep{Christensen:2001cr}. We calculate the posterior probability density function, $p(\bm{\theta}|\bm{d}(t),H)$, for the set of parameters $\bm{\theta}$ for the GW model, $H$, given the LIGO Hanford, Livingston, and Virgo GW data $\bm{d}(t)$:
\begin{equation}
p(\bm{\theta}|\bm{d}(t),H) = \frac{p(\bm{\theta}|H) p(\bm{d}(t)|\bm{\theta},H)}{p(\bm{d}(t)|H)}, 
\label{eq:postpdf}
\end{equation}
where $\bm{\theta}$ is the vector of the gravitational waveform parameters. The prior, $p(\bm{\theta}|H)$, is the set of assumed prior probability distributions for the waveform parameters. The likelihood $p(\bm{d}(t)|\bm{\theta},H)$ assumes a Gaussian model of detector noise and depends upon the noise-weighted inner product between the gravitational waveform and the GW detector data $\bm{d}(t)$~\citep{Finn:2000hj,Rover:2006bb}. Marginalization of the likelihood to obtain the posterior probabilities is performed using Markov Chain Monte Carlo (MCMC) techniques. Our implementation used the \textit{PyCBC Inference} software package \citep{pycbc-inference,pycbc-software} and the parallel-tempered \textit{emcee} sampler \citep{emcee}. 

The MCMC is performed over the detector-frame chirp mass of the binary $\mathcal{M}^\mathrm{det}$, the mass ratio $q = m_1/m_2, m_1 \ge m_2$, the component spins $\chi_{1,2}$, the time of coalescence $t_c$, the phase of coalescence $\phi_c$, the GW polarization angle $\psi$, the inclination angle of the binary $\iota$, R.A. and decl. of the binary, and the luminosity distance $d_L$. 

We assume a uniform prior distribution on the binary component masses, $m_{1,2}\in [1.0, 2.0]~M_{\odot}$, transformed to $\mathcal{M}^\mathrm{det}$ and $q$ with a cut on the detector-frame chirp mass $1.1876 \le \mathcal{M}^\mathrm{det} \le 1.2076$. We assume a uniform prior on the dimensionless angular momentum of each neutron star, $\chi_{1,2}\in [-0.05, 0.05]$ \citep{Brown:2012qf}. The prior on $t_c$ is uniform in the GPS time interval $[1187008882.3434,1187008882.5434]$. We assume a uniform prior between $0$ and $2\pi$ for $\phi_c$ and $\psi$. We incorporate EM information through fixing the R.A. and decl. of GW170817 and through the prior probability distribution on the luminosity distance $p(d_L|H)$. We run the MCMC with two prior distributions on the inclination angle $\iota$: a prior uniform in $\cos \iota$, and a prior uniform in $\iota$ to explore the posterior distribution for small viewing angles.

We use GW strain data from the Advanced LIGO and Virgo detectors for the GW170817 event, made available through the LIGO Open Science Center (LOSC) \citep{Vallisneri:2014vxa}. The \texttt{LOSC\_CLN\_16\_V1} data that we use here include a post-processing noise subtraction performed by the LIGO/Virgo Collaboration \citep{gw170817-losc,gw170817-noise}. The LOSC documentation states that these data have been truncated to remove tapering effects due to the cleaning process, however the LOSC data shows evidence of tapering after GPS time $1187008900$ in the LIGO Hanford detector. To avoid any contamination of our results we do not use any data after GPS time $1187008891$.  

We high-pass the GW data using an eighth-order Butterworth filter that has an attenuation of $0.1$ at 15~Hz. The filter is applied forward and backward to preserve the phase of the data. A low-pass (anti-aliasing) finite impulse response filter is applied prior to resampling the data. The data is decimated to a sample rate of 4096~Hz for the analysis. To estimate the detector's noise power spectral density (PSD) for computing the GW likelihood, we use Welch's method with 16-second Hann-windowed segments (overlapped by 8~s) taken from GPS time $1187007048$ to $1187008680$. The PSD estimate is truncated to 8~s length in the time domain using the method described in \cite{Allen:2005fk}. The GW data $\vec{d}(t)$ used in the likelihood is taken from the interval $1187008763$ to $1187008891$. The GW likelihood is evaluated from a low-frequency cutoff of 25~Hz to the Nyquist frequency of 2048~Hz. 

The waveform model $H$ is the restricted TaylorF2 post-Newtonian (pN) aligned-spin waveform model. We use the LIGO Algorithm Library implementation \citep{lal} accurate to 3.5 pN order in orbital phase \citep{Buonanno:2009zt}, 2.0 pN order in spin--spin, quadrupole--monopole and self-spin interactions \citep{Mikoczi:2005dn,Arun:2008kb}, and 3.5 pN order in spin--orbit interactions \citep{Bohe:2013cla}. The waveforms are terminated at twice the orbital frequency of a test particle at the innermost stable circular orbit of a Schwarzschild black hole of mass $M = m_1 + m_2$. We neglect matter effects in the waveforms as we find that their effect is significantly smaller than the statistical errors on our measurement of $d_L$ and $\iota$. 

To measure the systematic effect of calibration uncertainties we use the 68\% occurrence, 1$\sigma$ calibration uncertainty bounds for LIGO/Virgo's second observing run as detailed in \cite{Cahillane:2017vkb}. We adjusted the GW strain to the extreme cases of calibration error in amplitude and phase to determine the systematic effects on parameter measurement. The strain adjustment was done according to 
\begin{equation}
\tilde{d'}(f) = \left(1+\frac{\delta R(f)}{R(f)}\right) \ \tilde{d}(f)
\end{equation}
where $\tilde{d}(f)$ is the frequency-domain GW strain data, $\delta R/R$ is the relative response function error (in amplitude and phase), and $\tilde{d'}(f)$ is the resulting adjusted strain data \citep{Viets:2017yvy}.

\section{Results} \label{sec:results}

\begin{figure}[ht]
\includegraphics[width=0.45\textwidth]{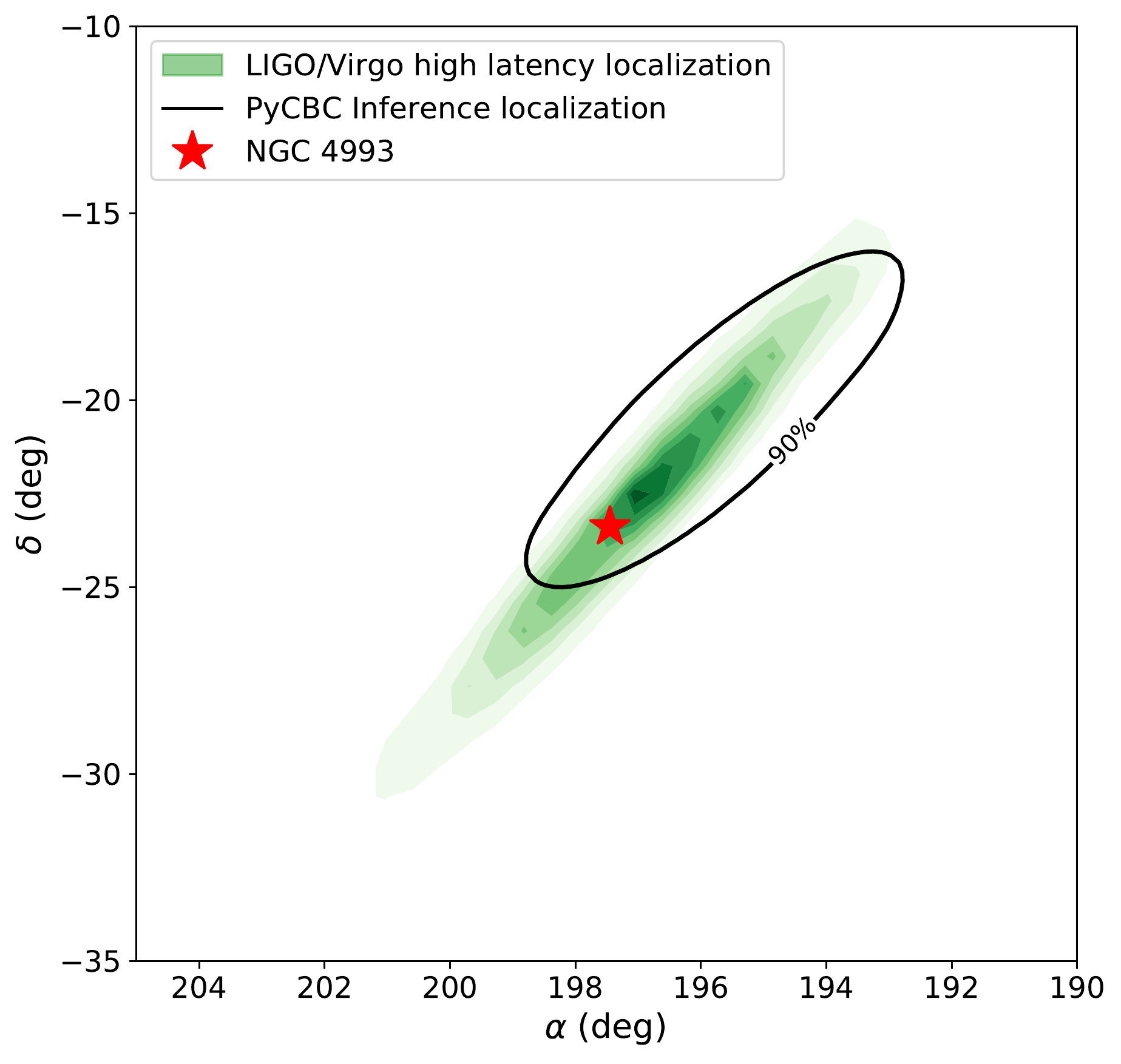}
\caption{Sky localizations of GW170817 for LIGO/Virgo (green shaded region) and from our analysis using only GW source information (black contour). Both localizations are 90\% confidence regions, while the LIGO/Virgo region shows contours at each 10\% threshold. The location of NGC\,4993 is marked as a red star.}
\label{fig:skyloc}
\end{figure}

As a check on our analysis, we first estimate the parameters of GW170817 using priors that do not assume any information about the source from EM observations.  We allow the R.A. and decl. to vary uniformly over the entire sky, and the distance to vary in a wide uniform-in-volume distribution of $[5, 80]$ Mpc. Our analysis localized the source to a region of $\approx 23$~deg$^{2}$ at 90\% confidence, shown in Figure~\ref{fig:skyloc}. Our sky localization encloses the location of NGC\,4993 (e.g., \citealt{Soares-Santos:2017lru}) and agrees well with the localization region of \cite{TheLIGOScientific:2017qsa}.

We then fix the sky location of GW170817 to R.A. = $197.450374^\circ$, decl. = $-23.381495^\circ$ \citep{Soares-Santos:2017lru} and remove these parameters from our parameter estimation.  Fixing the sky location of GW170817 has virtually no impact on the inclination measurement, in agreement with previous studies that have explored this correlation \citep{Seto:2005xb,Arun:2014ysa}. Finally, we set the prior probability distribution on the luminosity distance $p(d_L|H)$ to a Gaussian distribution centered on $40.7$ Mpc with a standard deviation of $2.36$ Mpc, corresponding to the measured distance and quadrature sum of statistical and systematic errors reported in \cite{Cantiello:2018ffy}. Here we have assumed a Gaussian distribution on this distance measurement, which we deem valid for a measurement of this precision and for the purpose of exploring upper and lower bounds.

\begin{figure*}[ht]
\includegraphics[width=\textwidth]{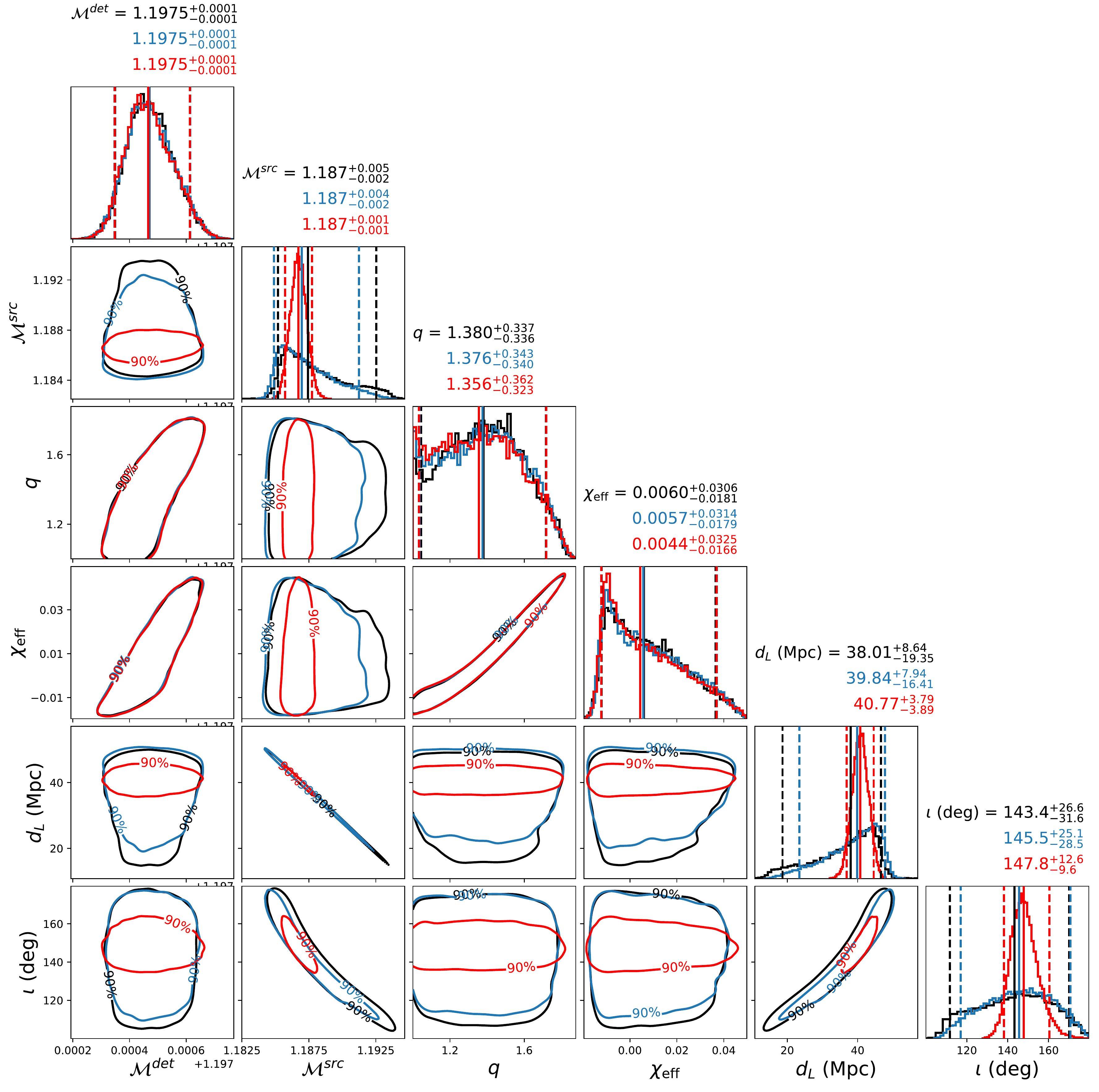}
\caption{Comparison of posterior probability distributions without and with combined EM information. The black contours show the results for using only the GW signal, the blue contours show the results for the fixed sky location of NGC\,4993, and the red contours show the results for both the fixed sky location and a Gaussian prior on distance of $40.7\pm 2.36$ Mpc from \cite{Cantiello:2018ffy}. These analyses used a prior on inclination angle that is uniform in $\cos \iota$. For each parameter, we quote the median value and the 90\% credible interval (shown with vertical solid and dashed lines, respectively, on the posterior plot of each parameter). The EM information on the distance measurement greatly improves the precision with which we measure the inclination angle (and significantly reduces the uncertainty on the source-frame chirp mass).}
\label{fig:posteriors}
\end{figure*}

Using the EM observations as the prior on the luminosity distance results in significantly narrower posteriors on inclination angle and source-frame chirp mass $\mathcal{M} = (m_1 m_2)^{3/5}/(m_1 + m_2)^{1/5}$ shown in Figure~\ref{fig:posteriors}. The improved chirp mass measurement is due to the reduced error on $d_L$, as the $d_L$ posterior samples are used to convert from the measured detector-frame chirp mass $\mathcal{M}^\mathrm{det}$ to the source-frame chirp mass $\mathcal{M}^\mathrm{src}$ \citep{Schutz:1986gp,Finn:1992xs}. However, the improved precision on distance has no effect on our measurements of the component masses or spins, because at leading order the mass ratio $q = m_1/m_2$ and spin parameter $\chi_\mathrm{eff}$ \citep{Cutler:1994ys} are not correlated with distance. With the EM observations as the prior on $d_L$ and a prior on the inclination angle uniform in $\cos\iota$, we find that the viewing angle is $\Theta = 32^{+10}_{-13}$ degrees (90\% confidence).

Errors in the calibration of the GW detectors can cause errors in the measured amplitude of the GW signal and hence in the inclination angle of the binary. We treat this as a systematic error, which we measure by shifting the amplitude calibration of the LIGO and Virgo detectors by the 1$\sigma$ uncertainty bounds for LIGO/Virgo's second observing run \citep{Cahillane:2017vkb}. We find that shifting the calibration to its most sensitive and least sensitive extremes results in a $\pm 1.7^\circ$ shift in the peak of the viewing angle when using a prior on inclination angle that is uniform in $\cos \iota$. We quote this shift as the systematic error on our measurement. Changing the phase error of the calibration within the bounds reported in \cite{Cahillane:2017vkb} produces a negligible effect on the inclination angle.

\begin{figure*}[ht]
\includegraphics[width=\textwidth]{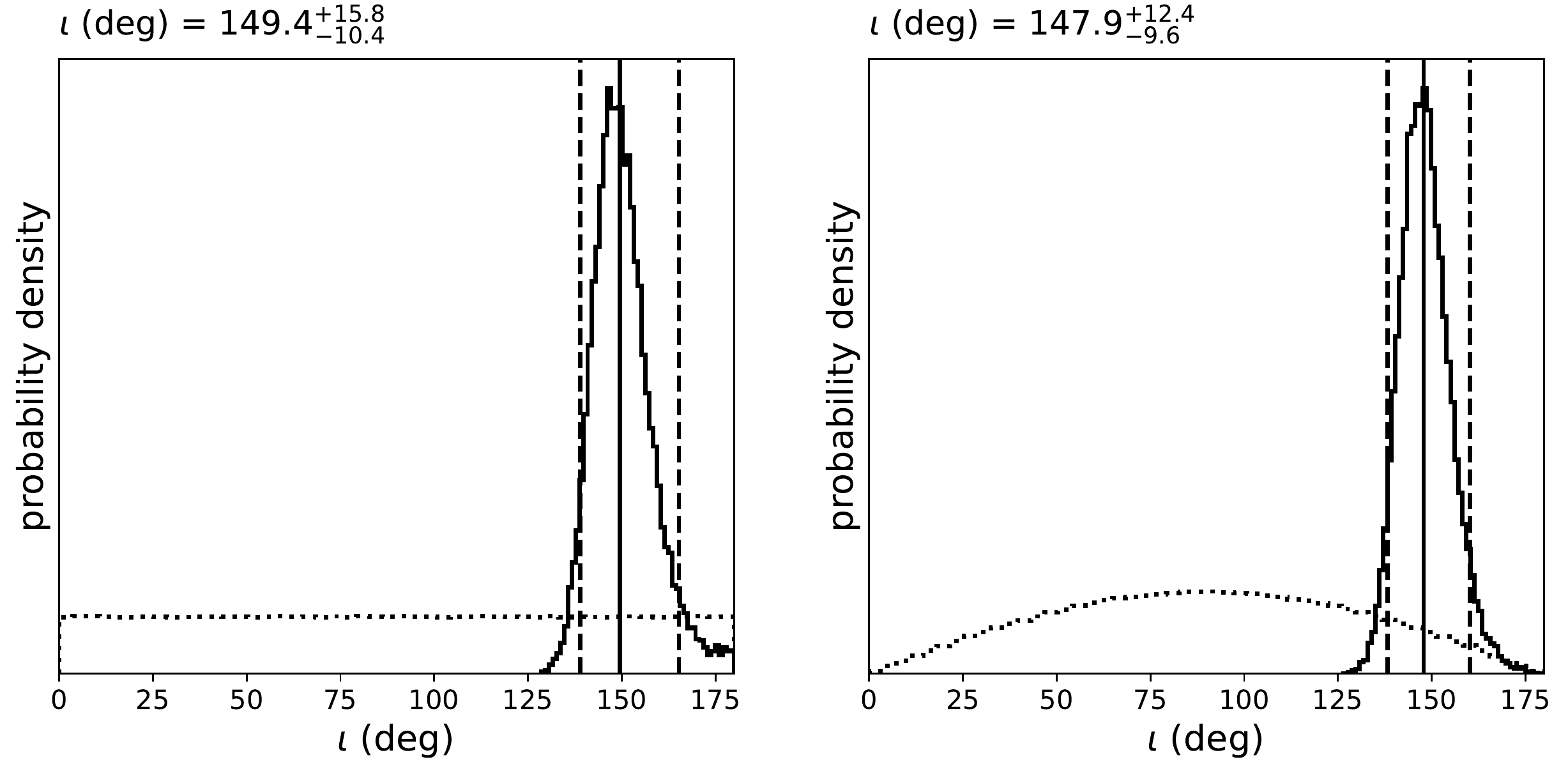}
\caption{Inclination angle posteriors (solid lines) plotted against their prior (dotted lines) for two choices of prior: uniform in $\iota$ (left), and uniform in $\cos \iota$ (right). We quote the median value and the 90\% credible interval for $\iota$ in each posterior (shown with vertical lines). The prior uniform in $\cos \iota$ is the prior used by the LIGO/Virgo analysis. The uniform prior does not bias measurement away from angles approaching $180^\circ$, so these results suggest that our likelihood is informative close to $\iota = 180^\circ$ and that we can place a conservative lower bound on the viewing angle $\Theta \ge 13^\circ$ (90\% confidence).}
\label{fig:inc}
\end{figure*}

A prior uniform in $\cos\iota$ goes to zero as the viewing angle approaches face on (or face off), so we repeat our analysis using a prior uniform in $\iota$. Figure~\ref{fig:inc} shows a comparison between the prior and the posterior distributions on inclination angle for each choice of prior. The result using a prior uniform in $\iota$ excludes viewing angles $\Theta \le 14.8^\circ$ at 90\% confidence, suggesting that our likelihood is indeed informative at small viewing angles and the lack of posterior support is not due to the prior uniform in $\cos\iota$ vanishing for small angles. Including the systematic error from calibration uncertainty, we set a conservative constraint of $\Theta \ge 13^\circ$ at 90\% confidence. This is consistent with the 10-day interval between the merger and the first observation of X-ray afterglow \citep{Troja:2017nqp}, which suggests that the GRB is not beamed at the Earth \citep{Guidorzi:2017ogy}.

\section{Discussion} \label{sec:disc}

Our joint GW--EM analysis of GW170817 used the GW observations along with sky location and a prior on the distance from direct measurement of these parameters from EM observations of NGC\,4993. Our 90\% confidence region on the viewing angle, $\Theta = 32^{+10}_{-13}\, \pm 1.7$ degrees (statistical and systematic errors), is significantly narrower than the inference made by GW observations alone, by about a factor of $2.6$.  It extends well above the $\Theta < 28^\circ$ bound of \cite{Mandel:2017fwk}, which was based on an assumed Hubble flow velocity for NGC\,4993. The precise distance measurement from \cite{Cantiello:2018ffy} also allows us to place a 90\% confidence lower bound on $\Theta$ that is substantially higher than the 68\% confidence lower bound, $\Theta > 10^\circ$, reported by \cite{Mandel:2017fwk}. 

Our improved constraint on $\Theta$ has implications for models of the prompt $\gamma$-ray and radio/X-ray afterglow emission from GW170817.  For example, our inferred value is in good agreement with the structured jet models of \cite{Lazzati:2017zsj}, which favor a viewing angle of $\approx 33^\circ$, and \cite{Margutti:2018xqd}, which favor a viewing angle of $\approx 20^\circ$.  While we do not yet know from a single event if the ejecta components that dominate the early UV/optical/near-infrared emission are significantly asymmetric, our constraint on $\Theta$ for GW170817 and future mergers will serve to shed light on the ejecta structure (e.g., spherical vs.~polar vs.~equatorial).

\acknowledgments

We thank Stefan Ballmer and Steven Reyes for useful discussions. We particularly thank Collin Capano and Alexander Nitz, who made significant contributions to the development of the PyCBC Inference software used in this Letter. This work was supported by U.S. National Science Foundation awards PHY-1404395 (D.A.B., C.M.B.), PHY-1707954 (D.A.B., S.D., D.F.) and AST-1714498 (E.B.). Computational work was supported by Syracuse University and National Science Foundation award OAC-1541396. D.A.B., E.B., and S.D. acknowledge the Kavli Institute for Theoretical Physics which is supported by the National Science Foundation award PHY-1748958.
All software used in this analysis is open source and available from \url{https://github.com/gwastro/pycbc}. Full posterior data samples from the MCMC are available at \url{https://github.com/sugwg/gw170817-inclination-angle}. The gravitational-wave data used in this work was obtained from the LIGO Open Science Center at \url{https://losc.ligo.org}. LOSC is a service of LIGO Laboratory, the LIGO Scientific Collaboration and the Virgo Collaboration. LIGO is funded by the National Science Foundation. Virgo is funded by the French Centre National de Recherche Scientifique (CNRS), the Italian Istituto Nazionale della Fisica Nucleare (INFN) and the Dutch Nikhef, with contributions by Polish and Hungarian institutes.

\bibliography{references}



\end{document}